\documentclass[%
aip,
pop,
amsmath,amssymb,
reprint,%
]{revtex4-1}

\usepackage{graphicx}% Include figure files
\usepackage{dcolumn}% Align table columns on decimal point
\usepackage{bm}% bold math
\usepackage{color}
\usepackage{times}
\usepackage[utf8]{inputenc}
\usepackage[T1]{fontenc}
\usepackage{mathptmx}
\usepackage{gensymb}
\usepackage{color}
\usepackage{amsmath}
\usepackage{url}
\usepackage{multirow}

\begin{document}

\title{Machine learning in the study of phase transition of two-dimensional complex plasmas}% Force line breaks with \\
%\thanks{Footnote to title of article.}

\author{He Huang}
 \affiliation{College of Science, Donghua University, 201620 Shanghai, PR China}
\author{Vladimir Nosenko}
 \affiliation{Institut f\"ur Materialphysik im Weltraum, Deutsches Zentrum f\"ur Luft- und Raumfahrt (DLR), 51147 Cologne, Germany}
\author{Han-Xiao Huang-Fu}
 \affiliation{College of Science, Donghua University, 201620 Shanghai, PR China}
\author{Hubertus M. Thomas}
 \affiliation{Institut f\"ur Materialphysik im Weltraum, Deutsches Zentrum f\"ur Luft- und Raumfahrt (DLR), 51147 Cologne, Germany}
\author{Cheng-Ran Du}
 \email{chengran.du@dhu.edu.cn}
 \affiliation{College of Science, Donghua University, 201620 Shanghai, PR China}
 \affiliation{Member of Magnetic Confinement Fusion Research Centre, Ministry of Education, PR China}
 
\date{\today}% It is always \today, today,
             %  but any date may be explicitly specified
 
\begin{abstract}
Machine learning is applied to investigate the phase transition of two-dimensional complex plasmas. The Langevin dynamics simulation is employed to prepare particle suspensions in various thermodynamic states. Based on the resulted particle positions in two extreme conditions, bitmap images are synthesized and imported to a convolutional neural network (ConvNet) as training sample. As a result, a phase diagram is obtained. This trained ConvNet model can be directly applied to the sequence of the recorded images using video microscopy in the experiments to study the melting.
\end{abstract}

% PACS, the Physics and Astronomy
                             % Classification Scheme.
%Use showkeys class option if keyword
                              %display desired
\maketitle

\section{introduction}

A complex plasma is a weakly ionized gas containing micron-sized dust particles \cite{Merlino2004PT,Morfill2009RMP,Fortov2005PR}. The particles are negatively charged owing to the higher thermal velocities of electrons compared to ions \cite{Goree1994PSST}. The discovery of the plasma crystal in the laboratory inspired lasting interest \cite{Thomas1994PRL,ILin1994PRL}. Using video microscopy, one can experimentally study plenty of phenomena such as wave propagation \cite{Menzel2010prl,Heidemann2009prl}, self-organization \cite{Suetterlin2009prl,Schwabe2011prl},  phase separation \cite{Wysockiprl2010,Killer2016prl}, and slow dynamics \cite{Du2019prl} at single-particle level \cite{Schwabe2018MST}. The particles are illuminated by a laser sheet and are recorded as bright dots by a fast video camera equipped with bandpass filter. The structure of the complex plasmas can then be obtained by direct processes such as Fourier transformation of the recorded images \cite{Du2019prl} or identification of \textit{x-y} positions of individual particles using particle tracking algorithms.  

Phase diagram and phase transition in complex plasmas are research topics of great interest \cite{Melzer1996PRE,Knapek2007PRLa,Khrapak2011prl,Wang2020prl}. In the laboratory, monodisperse particles are suspended in a single layer above the lower electrode where the gravity is compensated by the electrostatic force in the sheath \cite{Feng2010PRL,Nunomura2005PRL}. The thermodynamic state of such two-dimensional (2D) complex plasma can be easily controlled by the experimental conditions, such as gas pressure and discharge power \cite{Thomas2018PPCF,Ishihara2007JPD}. Tuning experimental parameters leads to the variation of two dimensionless parameters: the coupling parameter $\Gamma$ and screening parameter $\kappa$. The former represents the ratio of interaction strength to the kinetic temperature, while the latter defines the relative scale of interparticle distance to the Debye screening length.

Melting of plasma crystal can also be induced by the localized disturbance such as shocks or shear flows driven by external field \cite{Knapek2007PRLb,Harila2012POP,Couedel2018PRE,Couedel2010PRL,Jaiswal2020melting,Melzer2000PRE,Nosenko2002PRL,Schwabe2017NJP}. Besides, the lateral wave of a fast moving particle above or below the plasma crystal lattice leads to heat transport. Kinetic energy is transferred to the lattice particles via collisions with the self-propelled extra particle, resulting in the melting of the crystal lattice \cite{Du2014PRE,Laut2017prl}. 
  
Recently, machine learning has become a widely used analysis technique in addressing physical problems, such as Gardner transition \cite{Li2021PNAS}, phase transitions \cite{Nieuwenburg2017NP}, crystal structures classification \cite{Ziletti2018NC}. Meanwhile, various machine learning methods have also been applied to the analysis in the complex plasma research. For example, Bayesian optimization framework was applied to perform a nonlinear response analysis in a complex plasma \cite{Ding2021MachineLearning}. Multilayer perceptron was used to classify fcc, bcc, and hcp structure in the three-dimensional plasma crystal, where features are defined based on the particle positions \cite{Dietz2017PRE}. Support vector machine was used to locate the interface in a binary complex plasma directly based on the recorded images in the experiments performed in the PK-3 Plus laboratory on board the International Space Station \cite{He2019JI}. Convolutional neural network was applied to reconstruct the three-dimensional (3D) positions of particles in a dense dust cloud in a dusty plasma under weightlessness from stereoscopic camera images \cite{Himpel2021}.

In this paper, we apply a machine learning method to investigate the phase transition in 2D complex plasmas based on the Langevin dynamics simulations. In Sec.~\ref{sec:method}, the numerical simulation is briefly introduced. The data process protocol and the machine learning model are described. In Sec.~\ref{sec:result}, the phase diagram is obtained using a convolutional neural network and compared with the ones obtained using other approaches. In Sec.~\ref{sec:application}, the trained model is directly applied to identify the change of the state of the complex plasma as it melts. Finally, a conclusion is drawn in Sec.~\ref{sec:conclusion}.

%==============================================================
\section{Method} 
\label{sec:method}

\begin{figure}[htbp]
	\centering
	\includegraphics[width=24pc]{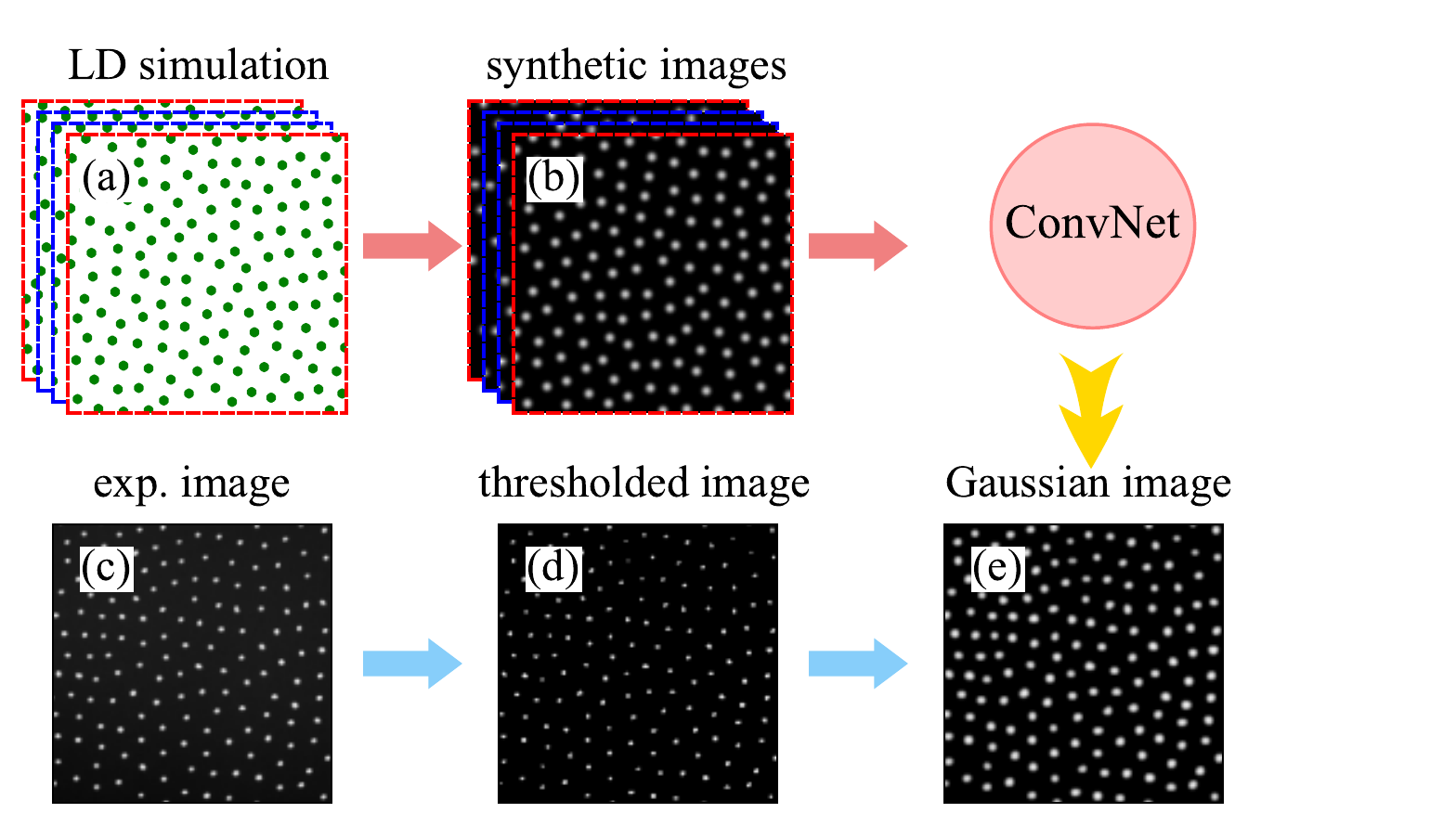}
	\caption{Scheme for applying machine learning in the phase transition of 2D complex plasmas. Training samples are prepared using Langevin dynamics simulations and fed to the ConvNet, a convolutional neural network (a,b). The trained model can be applied to identify the thermodynamic state of the complex plasma in the experiment, where the raw images of the experimental recordings are thresholded and filtered by a Gaussian kernel (c-e).}
	\label{fig:scheme}
\end{figure}
 
The Langevin dynamics simulation is applied to prepare complex plasmas in various thermodynamic states. For the purpose of model training and later application in the analysis of experiments, sequences of images are prepared based on the simulation results. The trained model can be used to identify the thermodynamic state of the experiments. The scheme is illustrated in Fig.~\ref{fig:scheme}.  

\subsection{Simulation}
\label{subsec:sim}

\begin{figure}[htbp]
	\centering
	\includegraphics[width=20pc]{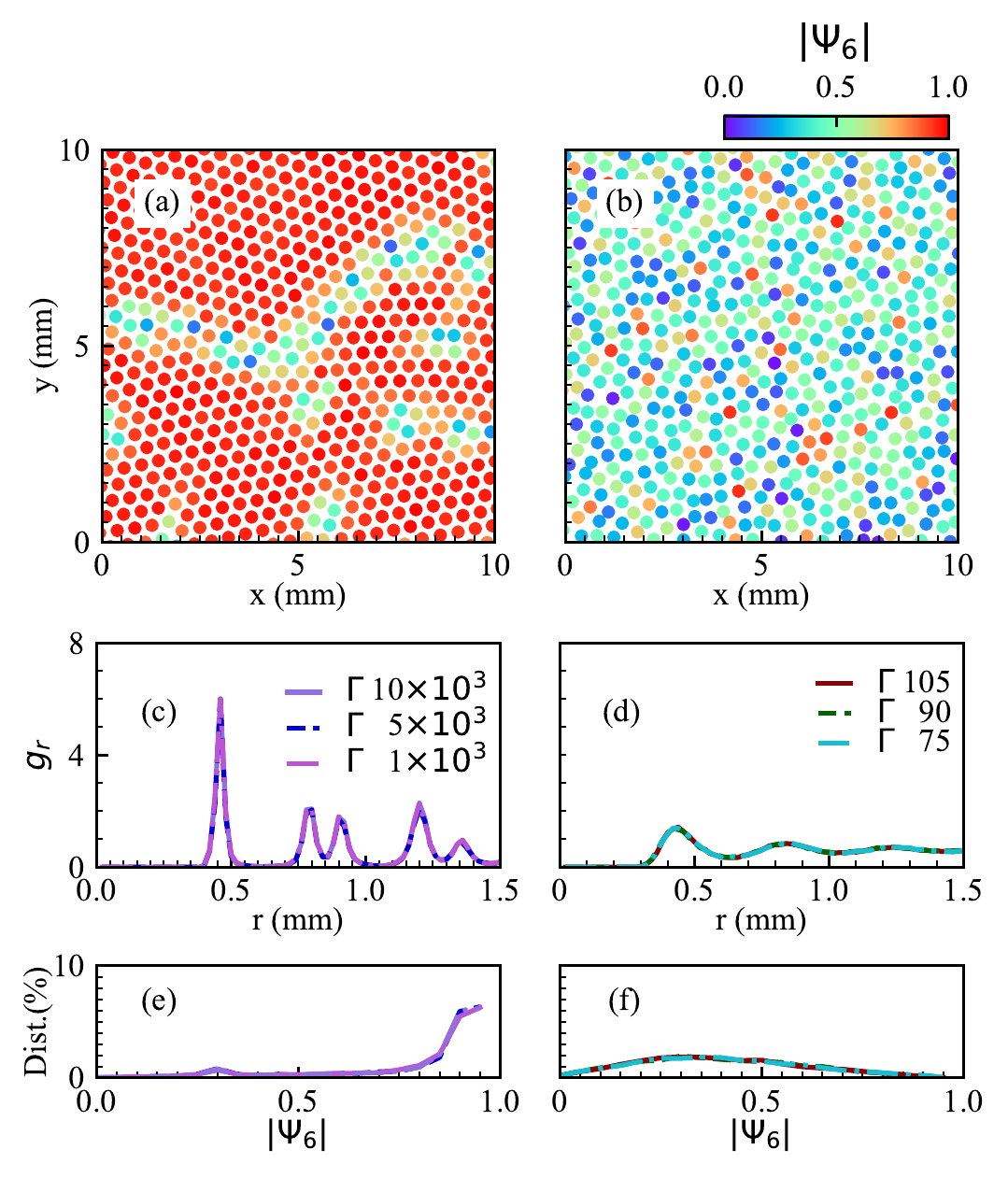}
	\caption{Comparison of the complex plasmas with different coupling parameters and $\lambda_D=400$~$\mu$m, corresponding to $\kappa=1.18$. The particle positions are color-coded with hexatic order parameter $|\Psi_6|$ in the simulation for $\Gamma=10000$ (a) and $\Gamma=100$ (b). The pair correlation function $g_r$ and the distribution of $|\Psi_{6}|$ (in percentage of particles) are shown for plasma crystals (c,e) and liquids (d,f) with various $\Gamma$ well above and below the melting point. }
	\label{fig:snapshots}
\end{figure}

 \begin{figure*}[bthp]
	\centering
	\includegraphics[scale=1.1, bb=0 30 800 130]{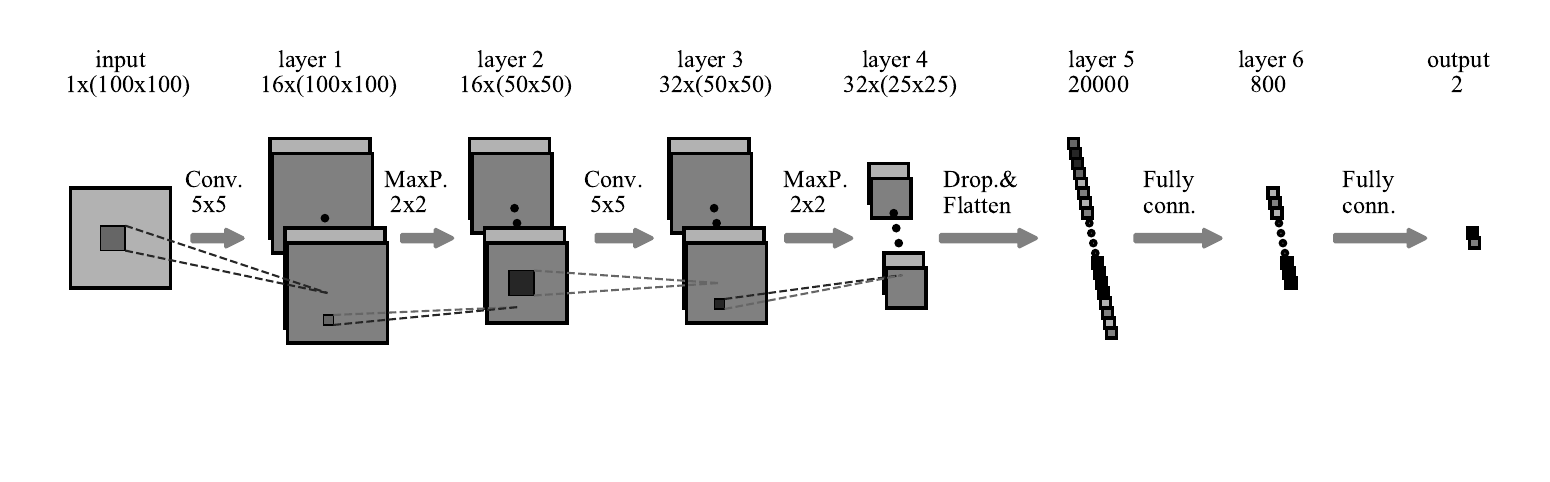}
	\caption{Architecture of the convolutional neural network (ConvNet) used for the classification. Two $5 \times 5$ kernel 2D convolution layers are included. Each convolutional layer is followed by a max-pooling layer. Before the fully connected layers, a Gaussian Dropout layer is added to prevent overfit. The last two layers are fully connected layers, to achieve binary classification.}
	\label{fig:architecture}
\end{figure*} 

The Langevin dynamics simulation is employed to prepare complex plasmas in different thermodynamic states. The equation of motion including the damping from the neutral gas and Brownian motion of microparticles is given by:
\begin{equation}
	\label{eq:motion}
	m_{i}\ddot{\bm{r}}_{i} +m_{i}\nu_{i}\dot{\bm{r}}_{i}=-\sum_{j\ne i}\bigtriangledown\phi_{ij} +\bm{L}_{i},         
\end{equation}
where $\bm{r}_{i}$ is the 2D position of particle $i$, $m$ is the particle mass, $\nu$ is the damping rate. The Langevin force $\bm{L}_{i}$ is defined by $\left \langle  \bm{L}_{i} \right \rangle=0$ and $\left \langle \bm{ L}_{i}(t)\bm{L}_{j}(t+\tau) \right \rangle = 2\nu m k_b T \delta_{i,j}\delta(\tau)\bm{I} $, where $k_b$ is the Boltzmann constant, $T$ is the temperature of the heat bath, $\delta_{ij}$ is Kronecker delta, $\delta(\tau)$ is the delta function, and $\bm{I}$ is the unit matrix. In the simulation, we assume that the particles interact with each other via the Yukawa potential,
\begin{equation}
	\label{eq:yukawa}
	\phi_{ij} =  \dfrac{Q_{i}Q_j}{4\pi\epsilon_{0}r_{ij}}exp(-\dfrac{r_{ij}}{\lambda_D} ) ,
\end{equation}
where $\lambda_D$ is the Debye length,$Q_i$ is the charge of particle $i$ and $Q_j$ is the charge of a neighboring particle $j$, the interparticle distance is $r_{ij}$. Here, we select typical experimental parameters in the simulations. The mass is set to $3\times10^{-13}$~kg and the particle charge is set to $8000$~$e$. The damping rate is set to $0.9$~$s^{-1}$. The total number of particles in the simulation is $6400$. Periodic boundary conditions are used. 

Two crucial dimensionless parameters to describe the strongly coupled systems with Yukawa interaction are the coupling parameter $\Gamma$ and screening parameter $\kappa$. The former is defined as
\begin{equation}
	\label{eq:coupling}
	\Gamma = \dfrac{Q^{2}}{4\pi\epsilon_{0}\Delta k_b T},
\end{equation}
where $\Delta$ is the length scale of the distance to the nearest neighbors and can be calculated as $\Delta =(\pi n_{d})^{-1/2}$, and $n_{d}$ is the particle number density \cite{Melzer1996PRE}. The latter is defined as
 \begin{equation}
 	\label{eq:screening}
 	\kappa=\frac{\Delta}{\lambda_D}.
 \end{equation}
In the simulations, we set the kinetic temperature $T$ to tune the coupling parameter and the Debye length $\lambda_D$ to tune the screening length. In order to cover the typical parameter range in complex plasmas, the temperature ranges from $100$ to $70000$~K and the Debye length ranges from $150$ to $1800$~$\mu$m for $\Delta\sim460$~$\mu$m, corresponding to $30 \lesssim \Gamma \lesssim 22000$ and $0.25 \lesssim \kappa \lesssim 3$. As a result, complex plasmas in various thermodynamic states from liquid to crystal are prepared.

Snapshots of particle positions with two extreme coupling strengths are demonstrated in Fig.~\ref{fig:snapshots}(a,b). In order to quantify the local structure , we define the hexatic order parameter $\Psi_{6,i}$ as
\begin{equation}	
	\label{eq:order}
	\Psi_{6,i}  = \frac{1}{6}\sum_{k=1}^{6}{e^{j6\theta_{k}}}, 
\end{equation}
where we only consider six nearest neighbors and $\theta_{k}$ is the angle between $\bm{r}_{k}-\bm{r}_{i}$ and the $x$ axis. The color coding in the snapshots represents the order parameter $|\Psi_{6,i}|$ of particles in simulation. For $|\Psi_{6,i}|=1$, the particle $i$ is located in the center of  a perfect hexagon cell, while for  $|\Psi_{6,i}|=0$, the particle $i$ is in a completely disordered structure. 

On the one hand, for a strongly coupled 2D complex plasma, the particles self-organize in a triangular lattice with hexagonal symmetry. A chain of dislocations lies in the middle of the snapshot, as we see in Fig.~\ref{fig:snapshots}(a). We select three strong coupling situations ($\Gamma=1,5,10\times10^{3}$) where the temperature is well below the melting temperature $T_m$, and the pair correlation functions $g_r$ as well as the distributions of  $|\Psi_{6}| $ are rather similar. These simulation results can be labeled as plasma crystals. On the other hand, when the coupling parameter is small, the particle suspension is in a liquid state and does not exhibit ordered structure, as shown in Fig.~\ref{fig:snapshots}(b). For three selected temperatures well above the melting temperature, $g_r$ and the distribution of  $|\Psi_{6}| $ do not differ much. If the temperature further increases, the first peak value of $g_r$ and the averaged $|\Psi_6|$ may also decrease. However, the degree of the variation is much smaller than that close to the melting temperature and thus has marginal influence on the further analysis. The detailed structure of 2D liquid complex plasma is rather complicated and beyond the scope of this paper \cite{Ott2014PRE,Ott2015CPP,Castello2021PRE}.

\subsection{Data Preparation}
\label{subsec:data}

The particle positions at each time step in the Langevin dynamics simulation are transformed into a gray-scale bitmap image, resembling the images exported from the video recording in the experiments. In order to achieve this purpose, a few steps are necessary. First, the particle positions $\bm{r}$ in SI unit are transformed into positions $\bm{R}_{x,y}$ in pixels by a coefficient $\eta$, so that the interparticle distances (in pixels) appear comparable with those in the recorded images. Second, the positions are ceiled, corresponding to the indices of a matrix $\bm{M}$ representing a gray-scale image. This matrix reads
\begin{equation}
\label{eq:image}
\bm{M}_{i,j}=\left\{
\begin{array}{rcl}
255   &      & (i,j) \in \lceil \bm{R}_{x,y} \rceil \\
0     &      & (i,j) \notin \lceil \bm{R}_{x,y} \rceil
\end{array} \right. .
\end{equation}
Finally, a Gaussian filter is applied to the matrix, convolving binary image with a Gaussian kernel $G_0\left( x,y\right) = 1/ {2 \pi {\sigma}^2 } \cdot \exp[ -(x^2 + y^2 )/ 2{\sigma}^2 ]$ and creating a gray-scale synthetic image. Here, the variance is set as ${\sigma} = 9$. The synthetic images are shown in Fig.~\ref{fig:scheme}(b).

Similar procedures are applied to the images obtained in the experiments. Despite the fact that the particles in the recorded bitmap images exhibit Gaussian profiles \cite{Feng2007RSI}, certain deviations still exist, especially in the presence of overexposure. To mitigate the discrepancy while applying the method to the experiment analysis, we apply a threshold to the experimental images to remove noise, binarize the gray-scale bitmaps and apply the Gaussian filter to them, as illustrated in Fig.~\ref{fig:scheme}(c-e). These steps may not be necessary, but can improve the performance of the algorithm to some extent.

\subsection{Machine Learning} 
\label{subsec:ml}

ConvNet, also known as convolutional neural network, is a specific type of deep learning network, which has been widely used in image identification in the past years \cite{LeCun1989,Krizhevsky2012,lin2014network}. In this work, we apply ConvNet to investigate the phase behaviors of 2D complex plasmas. The architecture of our network is shown in Fig.~\ref{fig:architecture}, similar to Lenet5 \cite{Lecun1998}. It contains $2$ convolutional layers, $2$ max-pooling layers, $1$ Gaussian Dropout layer and $2$ fully connected layers.
 
Two $5 \times 5$ kernel 2D convolution layers are used. The convolutional layers extract increasingly high-level feature representations of the input and preserve their spatial relationship. The expression of the convolution can be written as:
\begin{equation}
 \bm{h}_{k}^{l}   =   \varphi( \sum_{j}\bm{W}_{j,k}^l\bm{h}_{j}^{l-1} + \bm{b}_k^l ) ,
\end{equation}
where $l$ is the layer index, $j$ is the index of input feature maps, $k$ is the index of output feature maps. The input $\bm{h}_{j}^{l-1}$ is the $j$th feature map at layer $l-1$, $\bm{h}_{k}^{l} $ is the $k$th feature map at layer $l$, $\bm{W}$ is the convolutional weight tensor and the value is random uniform. $\bm{b}$ is the bias term and the value is initialized to zero. $\varphi(\cdot)$ is the element-wise  nonlinearity function and we used the rectified linear unit (RELU) function \cite{2019Hinton}. The number of channels is set to $16$ for the first convolutional layer and $32$ for the second convolutional layer.
 
Each convolutional layer is followed by a max-pooling layer. The max-pooling layers only preserve the maximum value within a local receptive field and discard all other values. By applying max-pooling layers, we reduce the number of free parameters and introduce a small amount of translational invariance into the network. Before the fully connected layers, we use the Gaussian Dropout layer to prevent overfit \cite{Srivastava2014}. The last two layers are fully connected layers. The first fully connected layer flattens all of the feature maps after the max-pooling layer. The second fully connected layer works like a linear classifier. We choose RELU for the fully connected hidden layers.
 
The ConvNet is easy to deploy using open source platform for machine learning. Our ConvNet is based on TensorFlow \cite{abadi2016tensorflow} and Keras \cite{chollet2015keras}. The training sample contains $2000$ labeled gray-scale bitmap images with a size of $100 \times 100$~pixel$^2$.  The training was conducted using Adam optimization \cite{Adam2014} with batches of $100$ images for $20$ epochs with a learning rate $10^{-3}$ and categorical crossentropy as loss function.  We train the ConvNet on $80\%$ of the dataset, and use the remaining $20\%$ as test set. Almost $100\%$ accuracy is achieved on the training and test datasets. This means that our model can learn the samples and at the same time is capable of classifying the images which have not been seen before. 
 
 %==============================================================
\section{Phase diagram}
\label{sec:result}

\begin{figure}[htbp]
	\centering
	\includegraphics[width=20pc]{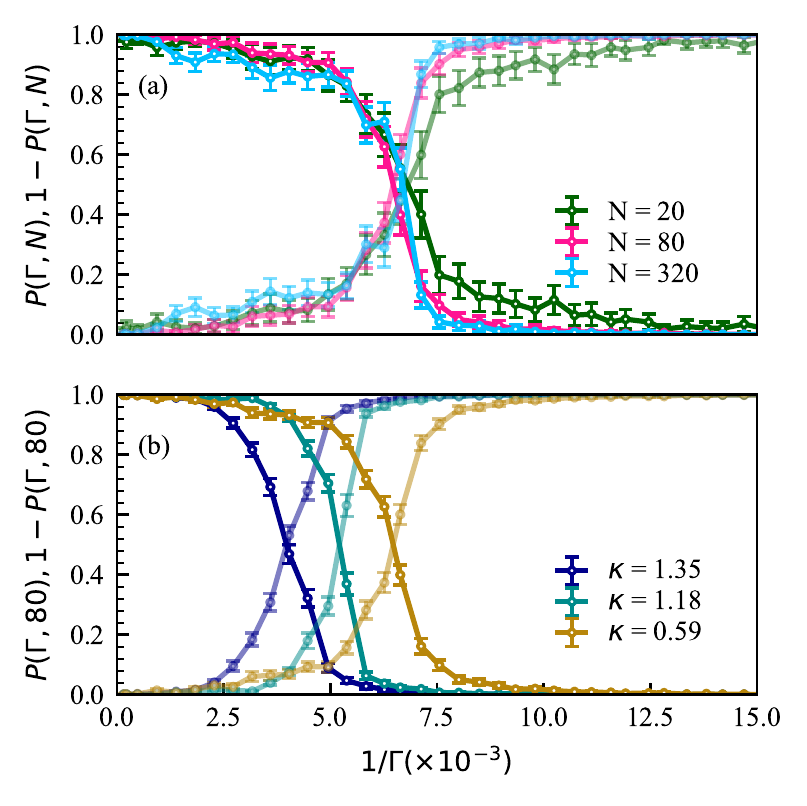}
	\caption{Probability $P(\Gamma,N)$ of complex plasma being classified as crystal (dark symbols) and probability $1-P(\Gamma,N)$ of being classified as liquid (light symbols), as function of coupling parameter $\Gamma$. (a) shows the dependence on the size of the sample image, while (b) shows the dependence on the screening parameter $\kappa$. }
	\label{fig:probability}
\end{figure}

Supervised machine learning method is applied to investigate the phase diagram of 2D complex plasmas. In the conventional supervised learning, the training samples are labeled based on some other criteria (usually not included in the traning sample), such as Lindemann measures, for example, in the investigation of phase behaviors of matter \cite{Chakravarty2007JCP}. This requires additional diagnostics of the particle motion while labeling the samples. In addition, the training data usually shall cover the whole range of thermodynamic states of interest. 

However, in this work, we try a different approach \cite{Nieuwenburg2017NP,Carrasquilla2017NP,Li2021PNAS}. Although the training samples still need to be labeled, as usually done in the supervised learning, they only include the thermodynamic states in two extreme scenarios. For high coupling ($\Gamma \gtrsim 1000$) the first peak value of $g_r$ is higher than $7$, while for low coupling ($\Gamma \lesssim 100$) the first peak value of $g_r$ is lower than $2$. The systems are in crystalline and liquid states, respectively. We do not have to provide training sample with a thermodynamic state near the critical value, to avoid ambiguity\footnote{We select extreme scenarios at different coupling parameter, train the model with selected training sample, and test its performance. It turns out that the selection of the exact $\Gamma$ has marginal impact to the model training as long as they are far beyond and below the critical value.}.

\begin{figure}[thbp]
	\centering
	\includegraphics[scale=0.7]{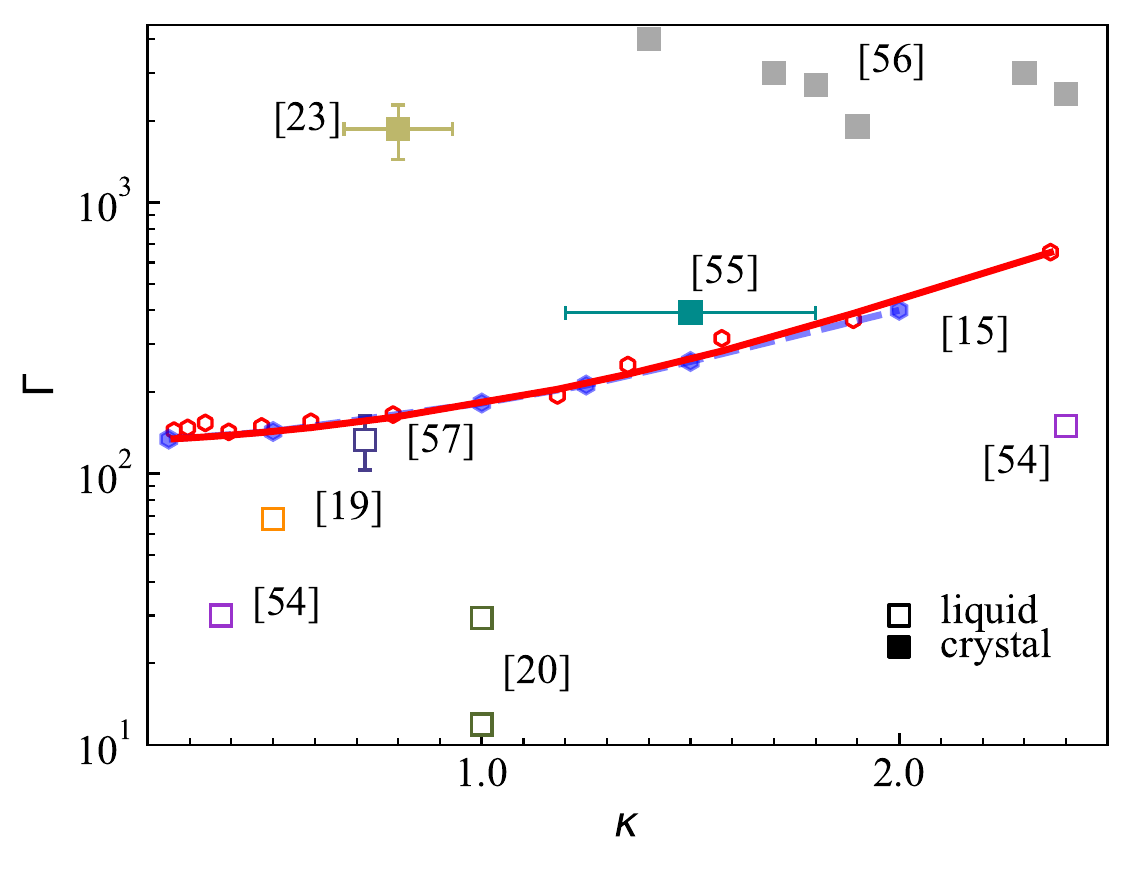}
	\caption{ Phase diagram obtained using ConvNet based on the Langevin dynamics (red hexagons), fitted by the analytical expression Eq.~\ref{eq:fit} (red line). Melting lines based on 2D-YOCP theory and on the numerical simulation with the criterion $\langle |\Psi_6| \rangle=0.45$ are shown by blue dashed line and blue solid hexagons, respectively. The experiments with plasma crystals and liquids are marked by solid and empty symbols, respectively. }
	\label{fig:phase}
\end{figure}

Once the training is completed, we apply this trained model to classify the synthetic images of 2D complex plasma resulting from the Langevin simulation. The simulation method has already been introduced in Sec.~\ref{subsec:sim}. We prepare particles with $T=100$~K, well below the melting temperature, in the simulation box and have the system relax. Then we slowly heat the system until it melts and further increase the temperature until $T=70000$~K. Synthetic bitmap images are generated every thousand K during the temperature increase. The simulation is repeated $20$ times with different initial conditions. The bitmap images are fed to the ConvNet model and result in the probability $P(\Gamma,N)$, that the image is classified as plasma crystal. Here, $N$ is the approximate number of particles included in the bitmap image sample. The cross of $P(\Gamma,N)$ and $1-P(\Gamma,N)$ provides the coupling strength of the phase transition $\Gamma_m$.

The dependence of $P(\Gamma,N)$ on the size of bitmap images is shown in Fig.~\ref{fig:probability}(a). The number of particles $N=20, 80, 320$ included in the bitmap image sample corresponds to the image sizes $50 \times 50,  100 \times 100, 200 \times 200$~pixel$^2$, respectively. The results show that for images including more than $20$ particles, the sample image size has marginal impact on the identification of $\Gamma_m$. In order to study the dependence of $\Gamma_m$ on the screening parameter, we also vary $\kappa$ by changing the Debye length $\lambda_D$ in the simulation. As shown in Fig.~\ref{fig:probability}(b), as the screening parameter $\kappa$ increases, the cross of $P(\Gamma,N)$ and $1-P(\Gamma,N)$ (denoting $1/\Gamma_m$) shifts to the left.

The phase diagram resulting from our ConvNet is shown as red hexagons in Fig.~\ref{fig:phase}. The melting line can be fitted by the analytical expression \cite{Hartmann2005PRE}
\begin{equation}
\label{eq:fit}
\Gamma_m\left( \kappa\right) = \frac{\Gamma_m^{OCP}  }{1 + a\kappa^2 +  b\kappa^3 + c\kappa^4 },
\end{equation}
where $\Gamma_m^{OCP}=131$ is an approximation for the 2D-OCP melting point \cite{Totsuji2004PRE,Vaulina2002PRE}. The fitting results $a=-0.401, b=0.132, c=0.0099$ show a fairly good agreement with the 2D-YOCP theory  \cite{Melzer1996PRE} (blue dashed line) and numerical simulations, where the melting line is found with the criterion $\langle |\Psi_6| \rangle=0.45$ \cite{Hartmann2005PRE} (blue solid hexagons), as well as other experiments listed in Tab.~\ref{tab:experiments}. 

 \begin{table}[htbp]
	\centering
	\caption{Dimensionless parameters $\Gamma$ and $\kappa$ and thermodynamic state of 2D complex plasmas in various experiments, corresponding to the symbols in Fig.~\ref{fig:phase}.}
	\label{tab:experiments}  
	\begin{tabular}{|c|c|c|c|}  
		\hline
		& & & \\[-6pt]
		Ref.					&$\Gamma$			&$\kappa$		&State \\  
		\hline
		& & & \\[-6pt] 	

		\cite{Hartmann2010PRL} 	&$392$				&$1.5\pm0.3$ 	&crystal \\
    	\cite{Knapek2007PRLb} 	&$1850\pm450$		&$0.8\pm0.13$ 	&crystal \\
        \multirow{2}{*}{\cite{Samsonov2000PRE}} & $3000,2500,2700$ & $2.3,2.4,1.8$ 	& \multirow{2}{*}{crystal} 	\\
        										& $1900,4000,3000$ & $1.9,1.4,1.7$	&							\\
    	\cite{Vorona2007} 		&$30,150$ 			&$0.375,2.4$	&liquid \\
    	\cite{Nunomura2005PRL} 	&$12,30$ 			&$1$ 			&liquid \\
    	\cite{Feng2010PRL} 		&$68$ 				&$0.5$ 			&liquid \\
        \cite{Haralson2016pop} 	&$92\sim155$		&$0.72$			&liquid \\
		
		\hline
	\end{tabular}
\end{table}

%==============================================================
\section{Application in Experiment}
\label{sec:application}

We apply our trained CovnNet to an experiment, to identify the state of the 2D complex plasma as it melts. The experiment was carried out in Gaseous Electronics Conference (GEC) rf reference cell \cite{Couedel2010PRL,YanFeng2008PRL}. Argon plasma was sustained using a capacitively coupled rf discharge at $13.56$~MHz. The input power was set at $20$~W. Melamine-formaldehyde(MF) spherical particles with a diameter of $7.17 \pm 0.07$~$\mu$m were levitated in the plasma sheath and illuminated by a horizontal laser sheet. The particle motion was recorded by a CMOS camera from the top. During the experiment, some disturbances were imposed by extra particles. 

 \begin{figure}[hbtp]
	\centering
	\includegraphics[width=18pc, bb=20 20 440 280]{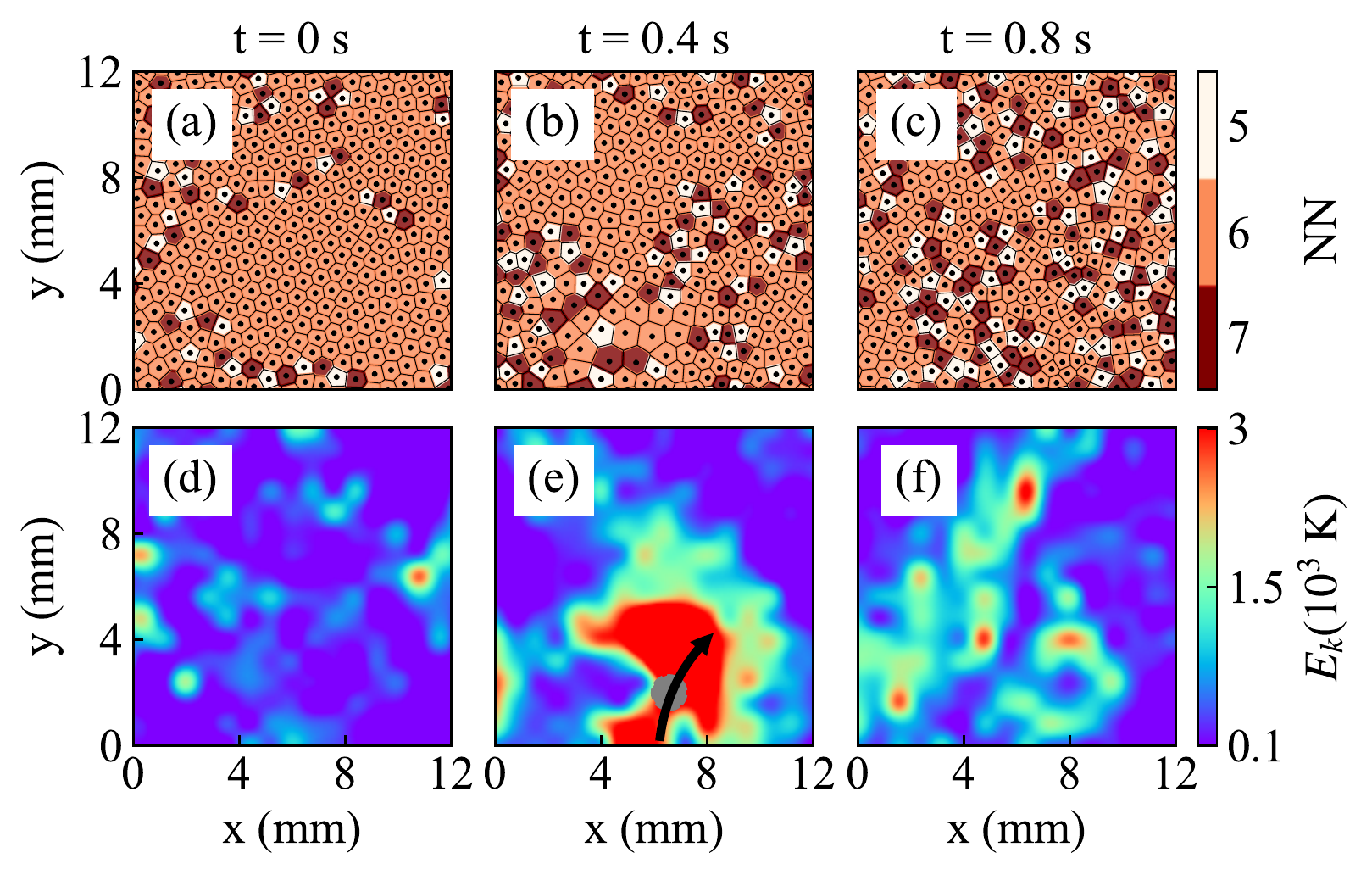}
	\caption{ Evolution of local structure and kinetic temperature, as an extra particle moves below the particle suspension. The structure is measured as the number of neighbors $NN$ using triangulation, whose value is $5$ for a five-fold defect and $7$ for a seven-fold defect. The extra particle is marked as gray circle in (e) and its motion is illustrated by an arrow.}
	\label{fig:melting}
\end{figure}

As an extra particle moved below the particle suspension with a high velocity, it created a strong disturbance in terms of Yukawa repulsion \cite{Samsonov2000PRE,Du2012EPL,Du2014PRE}. The interparticle distance of the particle suspension in the vicinity of the extra particle increases dramatically, and the ordered structure is destroyed, illustrated by the emergence of many defects locally, see Fig.~\ref{fig:melting}(b,e).  Heat is transported to the plasma crystal, local kinetic temperature spikes, and eventually the system melts, as shown in Fig.~\ref{fig:melting}(c,f). 

\begin{figure}[hbtp]
	\centering
	\includegraphics[scale=0.55]{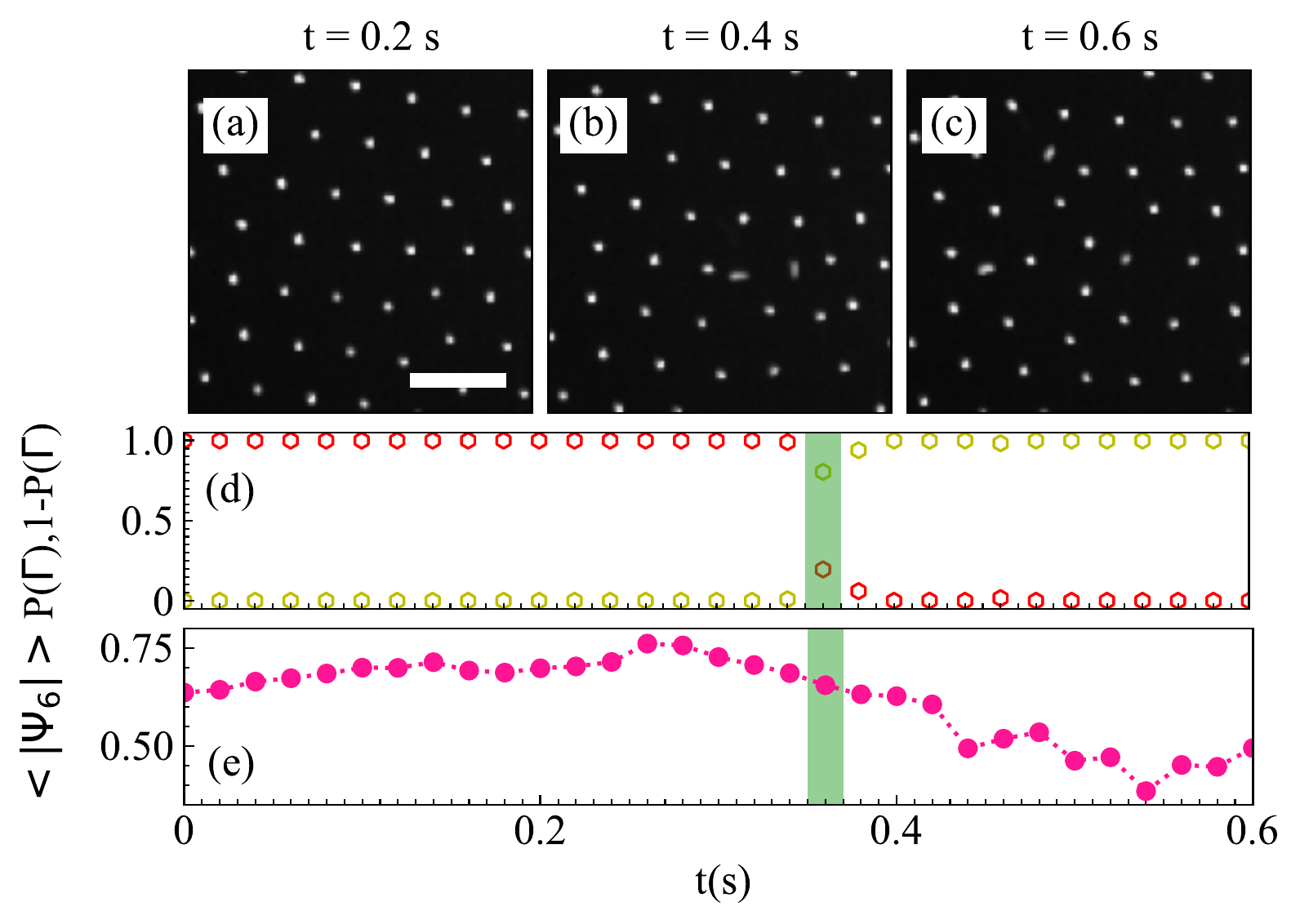}
	\caption{ Raw bitmap images as test sample for ConvNet (a-c), the evolution of test results (d) and the hexatic order parameter $\langle |\Psi_6| \rangle$. The probability $P(\Gamma)$ of being plasma crystal is shown by the red circles, while the probability $1-P(\Gamma)$ of being liquid is shown by the yellow circles in (d). The green strip in (d,e) highlights the moment when the extra particle moves into the region of interest. The scale bar corresponds to $1$~mm in (a).}
	\label{fig:application}
\end{figure}

The raw bitmap images [c.f. Fig.~\ref{fig:application}(a-c)] recorded in the experiment are processed according to the description in Sec.~\ref{subsec:data}. The processed images of the whole melting event are fed to the trained ConvNet and the output layer provides the probability $P$ of the particle suspension being crystalline. The evolution of $P$ and $\langle \Psi_6 \rangle$ are shown in Fig.~\ref{fig:application}(d,e), respectively. As the particle moves into the field of interest, $P$ drops instantaneously, signifying that the plasma crystal melts. This can also be seen as the substantial drop of $\langle \Psi_6 \rangle$ at the same time \cite{Hartmann2005PRE,Schweigert1999PRL}. By applying the machine learning method, a sharper contrast on the identification of the melting transition is achieved than that of the traditional method based on $\Psi_6$.

%==============================================================
\section{Conclusion} 
\label{sec:conclusion}

To conclude, we apply a machine learning method to study the phase transition of 2D complex plasmas. A convolutional neural network (ConvNet) is trained with the synthetic bitmap images based on the Langevin dynamics simulation, where definition of feature parameters is not needed. By training the model with the samples in two extreme scenarios in plasma crystal (very high $\Gamma$) and in liquid (very low $\Gamma$), a phase diagram is obtained, which agrees well with previous YOCP theory and numerical simulations. The method can be directly applied to the analysis of experiments, as demonstrated in this paper.

This method may be extended to investigate the phase transition in 3D complex plasmas, where particle tracking using video microscopy is challenging. As particle tracking is not necessary and thus the analysis is much faster than traditional methods, it can be possibly applied to the experiments on board the space station, where the automation of the experiment control based on the live diagnostics is desirable. We leave this for the future work.
  
\begin{acknowledgments}
The authors acknowledge the support from the National Natural Science Foundation of China (NSFC), Grant No.~$11975073$ and $21035003$. The students in Class $012781$ (Summer Semester $2021$) of the lecture "Machine Learning in Physics" have participated in developing the architecture of the machine learning model as their semester assignment. Their reports are taken as reference for this work. The authors acknowledge their contributions and inspirations. We thank M. Schwabe for the helpful discussions and C. Knapek for the valuable comments.
\end{acknowledgments}

%\bibliographystyle{unsrt}
%\bibliography{citation}

\begin{thebibliography}{10}
	
	\bibitem{Merlino2004PT}
	Robert~L. Merlino and John~A. Goree.
	\newblock Dusty plasmas in the laboratory, industry, and space.
	\newblock {\em Physics Today}, 57(7):32--38, 2004.
	
	\bibitem{Morfill2009RMP}
	Gregor~E. Morfill and Alexei~V. Ivlev.
	\newblock Complex plasmas: An interdisciplinary research field.
	\newblock {\em Rev. Mod. Phys.}, 81:1353--1404, Oct 2009.
	
	\bibitem{Fortov2005PR}
	V.E. Fortov, A.V. Ivlev, S.A. Khrapak, A.G. Khrapak, and G.E. Morfill.
	\newblock Complex (dusty) plasmas: Current status, open issues, perspectives.
	\newblock {\em Physics Reports}, 421:1 -- 103, 2005.
	
	\bibitem{Goree1994PSST}
	J~Goree.
	\newblock Charging of particles in a plasma.
	\newblock {\em Plasma Sources Science and Technology}, 3(3):400--406, aug 1994.
	
	\bibitem{Thomas1994PRL}
	H.~Thomas, G.~E. Morfill, V.~Demmel, J.~Goree, B.~Feuerbacher, and
	D.~M\"ohlmann.
	\newblock Plasma crystal: Coulomb crystallization in a dusty plasma.
	\newblock {\em Phys. Rev. Lett.}, 73:652--655, Aug 1994.
	
	\bibitem{ILin1994PRL}
	J.~H. Chu and Lin I.
	\newblock Direct observation of coulomb crystals and liquids in strongly
	coupled rf dusty plasmas.
	\newblock {\em Phys. Rev. Lett.}, 72:4009--4012, Jun 1994.
	
	\bibitem{Menzel2010prl}
	K.~O. Menzel, O.~Arp, and A.~Piel.
	\newblock Spatial frequency clustering in nonlinear dust-density waves.
	\newblock {\em Phys. Rev. Lett.}, 104:235002, Jun 2010.
	
	\bibitem{Heidemann2009prl}
	R.~Heidemann, S.~Zhdanov, R.~S\"utterlin, H.~M. Thomas, and G.~E. Morfill.
	\newblock Dissipative dark soliton in a complex plasma.
	\newblock {\em Phys. Rev. Lett.}, 102:135002, Mar 2009.
	
	\bibitem{Suetterlin2009prl}
	K.~R. S\"utterlin, A.~Wysocki, A.~V. Ivlev, C.~R\"ath, H.~M. Thomas,
	M.~Rubin-Zuzic, W.~J. Goedheer, V.~E. Fortov, A.~M. Lipaev, V.~I. Molotkov,
	O.~F. Petrov, G.~E. Morfill, and H.~L\"owen.
	\newblock Dynamics of lane formation in driven binary complex plasmas.
	\newblock {\em Phys. Rev. Lett.}, 102:085003, Feb 2009.
	
	\bibitem{Schwabe2011prl}
	M.~Schwabe, U.~Konopka, P.~Bandyopadhyay, and G.~E. Morfill.
	\newblock Pattern formation in a complex plasma in high magnetic fields.
	\newblock {\em Phys. Rev. Lett.}, 106:215004, May 2011.
	
	\bibitem{Wysockiprl2010}
	A.~Wysocki, C.~R\"ath, A.~V. Ivlev, K.~R. S\"utterlin, H.~M. Thomas,
	S.~Khrapak, S.~Zhdanov, V.~E. Fortov, A.~M. Lipaev, V.~I. Molotkov, O.~F.
	Petrov, H.~L\"owen, and G.~E. Morfill.
	\newblock Kinetics of fluid demixing in complex plasmas: Role of two-scale
	interactions.
	\newblock {\em Phys. Rev. Lett.}, 105:045001, Jul 2010.
	
	\bibitem{Killer2016prl}
	Carsten Killer, Tim Bockwoldt, Stefan Sch\"utt, Michael Himpel, Andr\'e Melzer,
	and Alexander Piel.
	\newblock Phase separation of binary charged particle systems with small size
	disparities using a dusty plasma.
	\newblock {\em Phys. Rev. Lett.}, 116:115002, Mar 2016.
	
	\bibitem{Du2019prl}
	Cheng-Ran Du, Vladimir Nosenko, Hubertus~M. Thomas, Yi-Fei Lin, Gregor~E.
	Morfill, and Alexei~V. Ivlev.
	\newblock Slow dynamics in a quasi-two-dimensional binary complex plasma.
	\newblock {\em Phys. Rev. Lett.}, 123:185002, Oct 2019.
	
	\bibitem{Schwabe2018MST}
	M.~Schwabe, C.-R. Du, P.~Huber, A.~M. Lipaev, V.~I. Molotkov, V.~N. Naumkin,
	S.~K. Zhdanov, D.~I. Zhukhovitskii, V.~E. Fortov, and H.~M. Thomas.
	\newblock Latest results on complex plasmas with the pk-3 plus laboratory on
	board the international space station.
	\newblock {\em Microgravity Science and Technology}, 30(5):581--589, Oct 2018.
	
	\bibitem{Melzer1996PRE}
	A.~Melzer, A.~Homann, and A.~Piel.
	\newblock Experimental investigation of the melting transition of the plasma
	crystal.
	\newblock {\em Phys. Rev. E}, 53:2757--2766, Mar 1996.
	
	\bibitem{Knapek2007PRLa}
	C.~A. Knapek, A.~V. Ivlev, B.~A. Klumov, G.~E. Morfill, and D.~Samsonov.
	\newblock Kinetic characterization of strongly coupled systems.
	\newblock {\em Phys. Rev. Lett.}, 98:015001, Jan 2007.
	
	\bibitem{Khrapak2011prl}
	S.~A. Khrapak, B.~A. Klumov, P.~Huber, V.~I. Molotkov, A.~M. Lipaev, V.~N.
	Naumkin, H.~M. Thomas, A.~V. Ivlev, G.~E. Morfill, O.~F. Petrov, V.~E.
	Fortov, Yu. Malentschenko, and S.~Volkov.
	\newblock Freezing and melting of 3d complex plasma structures under
	microgravity conditions driven by neutral gas pressure manipulation.
	\newblock {\em Phys. Rev. Lett.}, 106:205001, May 2011.
	
	\bibitem{Wang2020prl}
	Wen Wang, Hao-Wei Hu, and Lin I.
	\newblock Surface-induced layering of quenched 3d dusty plasma liquids:
	Micromotion and structural rearrangement.
	\newblock {\em Phys. Rev. Lett.}, 124:165001, Apr 2020.
	
	\bibitem{Feng2010PRL}
	Yan Feng, J.~Goree, and Bin Liu.
	\newblock Viscoelasticity of 2d liquids quantified in a dusty plasma
	experiment.
	\newblock {\em Phys. Rev. Lett.}, 105:025002, Jul 2010.
	
	\bibitem{Nunomura2005PRL}
	S.~Nunomura, S.~Zhdanov, D.~Samsonov, and G.~Morfill.
	\newblock Wave spectra in solid and liquid complex (dusty) plasmas.
	\newblock {\em Phys. Rev. Lett.}, 94:045001, Feb 2005.
	
	\bibitem{Thomas2018PPCF}
	H~M Thomas, M~Schwabe, M~Y Pustylnik, C~A Knapek, V~I Molotkov, A~M Lipaev, O~F
	Petrov, V~E Fortov, and S~A Khrapak.
	\newblock Complex plasma research on the international space station.
	\newblock {\em Plasma Physics and Controlled Fusion}, 61(1):014004, nov 2018.
	
	\bibitem{Ishihara2007JPD}
	Osamu Ishihara.
	\newblock Complex plasma: dusts in plasma.
	\newblock {\em Journal of Physics D: Applied Physics}, 40(8):R121--R147, apr
	2007.
	
	\bibitem{Knapek2007PRLb}
	C.~A. Knapek, D.~Samsonov, S.~Zhdanov, U.~Konopka, and G.~E. Morfill.
	\newblock Recrystallization of a 2d plasma crystal.
	\newblock {\em Phys. Rev. Lett.}, 98:015004, Jan 2007.
	
	\bibitem{Harila2012POP}
	S.~S. Harilal, G.~V. Miloshevsky, P.~K. Diwakar, N.~L. LaHaye, and
	A.~Hassanein.
	\newblock Experimental and computational study of complex shockwave dynamics in
	laser ablation plumes in argon atmosphere.
	\newblock {\em Physics of Plasmas}, 19(8):083504, 2012.
	
	\bibitem{Couedel2018PRE}
	L.~Cou\"edel, V.~Nosenko, M.~Rubin-Zuzic, S.~Zhdanov, Y.~Elskens, T.~Hall, and
	A.~V. Ivlev.
	\newblock Full melting of a two-dimensional complex plasma crystal triggered by
	localized pulsed laser heating.
	\newblock {\em Phys. Rev. E}, 97:043206, Apr 2018.
	
	\bibitem{Couedel2010PRL}
	L.~Cou\"edel, V.~Nosenko, A.~V. Ivlev, S.~K. Zhdanov, H.~M. Thomas, and G.~E.
	Morfill.
	\newblock Direct observation of mode-coupling instability in two-dimensional
	plasma crystals.
	\newblock {\em Phys. Rev. Lett.}, 104:195001, May 2010.
	
	\bibitem{Jaiswal2020melting}
	S.~Jaiswal and Ed~Thomas~Jr au2.
	\newblock Melting transition of two-dimensional complex plasma crystal in the
	dc glow discharge, 2020.
	
	\bibitem{Melzer2000PRE}
	A.~Melzer, S.~Nunomura, D.~Samsonov, Z.~W. Ma, and J.~Goree.
	\newblock Laser-excited mach cones in a dusty plasma crystal.
	\newblock {\em Phys. Rev. E}, 62:4162--4176, Sep 2000.
	
	\bibitem{Nosenko2002PRL}
	V.~Nosenko, J.~Goree, Z.~W. Ma, and A.~Piel.
	\newblock Observation of shear-wave mach cones in a 2d dusty-plasma crystal.
	\newblock {\em Phys. Rev. Lett.}, 88:135001, Mar 2002.
	
	\bibitem{Schwabe2017NJP}
	M~Schwabe, S~Zhdanov, T~Hagl, P~Huber, A~M Lipaev, V~I Molotkov, V~N Naumkin,
	M~Rubin-Zuzic, P~V Vinogradov, E~Zaehringer, V~E Fortov, and H~M Thomas.
	\newblock Observation of metallic sphere{\textendash}complex plasma
	interactions in microgravity.
	\newblock {\em New Journal of Physics}, 19(10):103019, oct 2017.
	
	\bibitem{Du2014PRE}
	Cheng-Ran Du, Vladimir Nosenko, Sergey Zhdanov, Hubertus~M. Thomas, and
	Gregor~E. Morfill.
	\newblock Channeling of particles and associated anomalous transport in a
	two-dimensional complex plasma crystal.
	\newblock {\em Phys. Rev. E}, 89:021101, Feb 2014.
	
	\bibitem{Laut2017prl}
	I.~Laut, C.~R\"ath, S.~K. Zhdanov, V.~Nosenko, G.~E. Morfill, and H.~M. Thomas.
	\newblock Wake-mediated propulsion of an upstream particle in two-dimensional
	plasma crystals.
	\newblock {\em Phys. Rev. Lett.}, 118:075002, Feb 2017.
	
	\bibitem{Li2021PNAS}
	Huaping Li, Yuliang Jin, Ying Jiang, and Jeff Z.~Y. Chen.
	\newblock Determining the nonequilibrium criticality of a gardner transition
	via a hybrid study of molecular simulations and machine learning.
	\newblock {\em Proceedings of the National Academy of Sciences}, 118(11), 2021.
	
	\bibitem{Nieuwenburg2017NP}
	Evert P. L. van Nieuwenburg, Ye-Hua Liu, and Sebastian D. Huber.
	\newblock Learning phase transitions by confusion.
	\newblock {\em Nature Physics}, 13(5):435–439, Feb 2017.
	
	\bibitem{Ziletti2018NC}
	Angelo Ziletti, Devinder Kumar, Matthias Scheffler, and Luca~M. Ghiringhelli.
	\newblock Insightful classification of crystal structures using deep learning.
	\newblock {\em Nature Communications}, 9(1), Jul 2018.
	
	\bibitem{Ding2021MachineLearning}
	Zhiyue Ding, Lorin~S Matthews, and Truell~W Hyde.
	\newblock A machine learning based bayesian optimization solution to non-linear
	responses in dusty plasmas.
	\newblock {\em Machine Learning: Science and Technology}, 2(3):035017, jun
	2021.
	
	\bibitem{Dietz2017PRE}
	C.~Dietz, T.~Kretz, and M.~H. Thoma.
	\newblock Machine-learning approach for local classification of crystalline
	structures in multiphase systems.
	\newblock {\em Phys. Rev. E}, 96:011301, Jul 2017.
	
	\bibitem{He2019JI}
	He~Huang, Mierk Schwabe, and Cheng-Ran Du.
	\newblock Identification of the interface in a binary complex plasma using
	machine learning.
	\newblock {\em Journal of Imaging}, 5(3), 2019.
	
	\bibitem{Himpel2021}
	Michael Himpel and Andr{\'{e}} Melzer.
	\newblock Fast 3d particle reconstruction using a convolutional neural network:
	application to dusty plasmas.
	\newblock {\em Machine Learning: Science and Technology}, 2(4):045019, sep
	2021.
	
	\bibitem{Ott2014PRE}
	T.~Ott, M.~Bonitz, L.~G. Stanton, and M.~S. Murillo.
	\newblock Coupling strength in coulomb and yukawa one-component plasmas.
	\newblock {\em Physics of Plasmas}, 21(11):113704, 2014.
	
	\bibitem{Ott2015CPP}
	T.~Ott and M.~Bonitz.
	\newblock First-principle results for the radial pair distribution function in
	strongly coupled one-component plasmas.
	\newblock {\em Contributions to Plasma Physics}, 55(2-3):243--253, 2015.
	
	\bibitem{Castello2021PRE}
	F.~Lucco Castello and P.~Tolias.
	\newblock Structure and thermodynamics of two-dimensional yukawa liquids.
	\newblock {\em Phys. Rev. E}, 103:063205, Jun 2021.
	
	\bibitem{Feng2007RSI}
	Y.~Feng, J.~Goree, and Bin Liu.
	\newblock Accurate particle position measurement from images.
	\newblock {\em Review of Scientific Instruments}, 78(5):053704, 2007.
	
	\bibitem{LeCun1989}
	Y.~LeCun, B.~Boser, J.~S. Denker, D.~Henderson, R.~E. Howard, W.~Hubbard, and
	L.~D. Jackel.
	\newblock Backpropagation applied to handwritten zip code recognition.
	\newblock {\em Neural Computation}, 1(4):541--551, 1989.
	
	\bibitem{Krizhevsky2012}
	Alex Krizhevsky, Ilya Sutskever, and Geoffrey~E. Hinton.
	\newblock Imagenet classification with deep convolutional neural networks.
	\newblock In {\em Proceedings of the 25th International Conference on Neural
		Information Processing Systems - Volume 1}, NIPS'12, page 1097–1105, Red
	Hook, NY, USA, 2012. Curran Associates Inc.
	
	\bibitem{lin2014network}
	Min Lin, Qiang Chen, and Shuicheng Yan.
	\newblock Network in network, 2014.
	
	\bibitem{Lecun1998}
	Y.~Lecun, L.~Bottou, Y.~Bengio, and P.~Haffner.
	\newblock Gradient-based learning applied to document recognition.
	\newblock {\em Proceedings of the IEEE}, 86(11):2278--2324, 1998.
	
	\bibitem{2019Hinton}
	Vinod Nair and Geoffrey~E. Hinton.
	\newblock Rectified linear units improve restricted boltzmann machines.
	\newblock In Johannes Fürnkranz and Thorsten Joachims, editors, {\em ICML},
	pages 807--814. Omnipress, 2010.
	
	\bibitem{Srivastava2014}
	Nitish Srivastava, Geoffrey Hinton, Alex Krizhevsky, Ilya Sutskever, and Ruslan
	Salakhutdinov.
	\newblock Dropout: A simple way to prevent neural networks from overfitting.
	\newblock {\em Journal of Machine Learning Research}, 15(56):1929--1958, 2014.
	
	\bibitem{abadi2016tensorflow}
	Martín Abadi, Ashish Agarwal, Paul Barham, Eugene Brevdo, Zhifeng Chen, Craig
	Citro, Greg~S. Corrado, Andy Davis, Jeffrey Dean, Matthieu Devin, Sanjay
	Ghemawat, Ian Goodfellow, Andrew Harp, Geoffrey Irving, Michael Isard,
	Yangqing Jia, Rafal Jozefowicz, Lukasz Kaiser, Manjunath Kudlur, Josh
	Levenberg, Dan Mane, Rajat Monga, Sherry Moore, Derek Murray, Chris Olah,
	Mike Schuster, Jonathon Shlens, Benoit Steiner, Ilya Sutskever, Kunal Talwar,
	Paul Tucker, Vincent Vanhoucke, Vijay Vasudevan, Fernanda Viegas, Oriol
	Vinyals, Pete Warden, Martin Wattenberg, Martin Wicke, Yuan Yu, and Xiaoqiang
	Zheng.
	\newblock Tensorflow: Large-scale machine learning on heterogeneous distributed
	systems, 2016.
	
	\bibitem{chollet2015keras}
	François Chollet et~al.
	\newblock Keras.
	\newblock \url{https://github.com/fchollet/keras}, 2015.
	
	\bibitem{Adam2014}
	Diederik~P. Kingma and Jimmy Ba.
	\newblock Adam: A method for stochastic optimization, 2014.
	\newblock cite arxiv:1412.6980Comment: Published as a conference paper at the
	3rd International Conference for Learning Representations, San Diego, 2015.
	
	\bibitem{Chakravarty2007JCP}
	Charusita Chakravarty, Pablo~G. Debenedetti, and Frank~H. Stillinger.
	\newblock Lindemann measures for the solid-liquid phase transition.
	\newblock {\em The Journal of Chemical Physics}, 126(20):204508, 2007.
	
	\bibitem{Carrasquilla2017NP}
	Juan Carrasquilla and Roger~G. Melko.
	\newblock Machine learning phases of matter.
	\newblock {\em Nature Physics}, 13(5):431–434, Feb 2017.
	
	\bibitem{Note1}
	We select extreme scenarios at different coupling parameter, train the model
	with selected training sample, and test its performance. It turns out that
	the selection of the exact $\Gamma $ has marginal impact to the model
	training as long as they are far beyond and below the critical value.
	
	\bibitem{Hartmann2005PRE}
	P.~Hartmann, G.~J. Kalman, Z.~Donk\'o, and K.~Kutasi.
	\newblock Equilibrium properties and phase diagram of two-dimensional yukawa
	systems.
	\newblock {\em Phys. Rev. E}, 72:026409, Aug 2005.
	
	\bibitem{Totsuji2004PRE}
	Hiroo Totsuji, M.~Sanusi~Liman, Chieko Totsuji, and Kenji Tsuruta.
	\newblock Thermodynamics of a two-dimensional yukawa fluid.
	\newblock {\em Phys. Rev. E}, 70:016405, Jul 2004.
	
	\bibitem{Vaulina2002PRE}
	O.~Vaulina, S.~Khrapak, and G.~Morfill.
	\newblock Universal scaling in complex (dusty) plasmas.
	\newblock {\em Phys. Rev. E}, 66:016404, Jul 2002.
	
	\bibitem{Hartmann2010PRL}
	Peter Hartmann, Angela Douglass, Jorge~C. Reyes, Lorin~S. Matthews, Truell~W.
	Hyde, Anik\'o Kov\'acs, and Zolt\'an Donk\'o.
	\newblock Crystallization dynamics of a single layer complex plasma.
	\newblock {\em Phys. Rev. Lett.}, 105:115004, Sep 2010.
	
	\bibitem{Samsonov2000PRE}
	D.~Samsonov, J.~Goree, H.~M. Thomas, and G.~E. Morfill.
	\newblock Mach cone shocks in a two-dimensional yukawa solid using a complex
	plasma.
	\newblock {\em Phys. Rev. E}, 61:5557--5572, May 2000.
	
	\bibitem{Vorona2007}
	N~A Vorona, A.~V. Gavrikov, A~S Ivanov, O~F Petrov, V~E Fortov, and I~A
	Shakhova.
	\newblock Viscosity of a dusty plasma liquid.
	\newblock {\em Journal of Experimental and Theoretical Physics}, 105(4), 10
	2007.
	
	\bibitem{Haralson2016pop}
	Zach Haralson and J.~Goree.
	\newblock Temperature dependence of viscosity in a two-dimensional dusty plasma
	without the effects of shear thinning.
	\newblock {\em Physics of Plasmas}, 23(9):093703, 2016.
	
	\bibitem{YanFeng2008PRL}
	Yan Feng, J.~Goree, and Bin Liu.
	\newblock Solid superheating observed in two-dimensional strongly coupled dusty
	plasma.
	\newblock {\em Phys. Rev. Lett.}, 100:205007, May 2008.
	
	\bibitem{Du2012EPL}
	C.-R. Du, V.~Nosenko, S.~Zhdanov, H.~M. Thomas, and G.~E. Morfill.
	\newblock Interaction of two-dimensional plasma crystals with upstream charged
	particles.
	\newblock {\em {EPL} (Europhysics Letters)}, 99(5):55001, sep 2012.
	
	\bibitem{Schweigert1999PRL}
	I.~V. Schweigert, V.~A. Schweigert, and F.~M. Peeters.
	\newblock Melting of the classical bilayer wigner crystal: Influence of lattice
	symmetry.
	\newblock {\em Phys. Rev. Lett.}, 82:5293--5296, Jun 1999.
	
\end{thebibliography}

\end{document}